\documentclass[aps,showpacs,twocolumn,letter]{revtex4}

\usepackage{graphicx}
\def\be{\begin{equation}}
\def\ee{\end{equation}}
\def\bea{\begin{eqnarray}}
\def\eea{\end{eqnarray}}
\begin{document}

\title{Pioneer anomaly? Gravitational pull due to the Kuiper belt}

\author{Jos\'e A. de Diego$^a$, Dar{\'\i}o N{\'u}{\~n}ez$^b$, Jes\'us
Zavala$^b$}
\affiliation{$^a$Instituto de Astronom\'\i a, \\
Universidad Nacional Aut{\'o}noma de M{\'e}xico\\
A. P. 70-262, 04510 M{\'e}xico, D. F., MEXICO \\
$^b$Instituto de Ciencias Nucleares, \\
Universidad Nacional Aut{\'o}noma de M{\'e}xico\\
A. P. 70-543, 04510 M{\'e}xico, D. F., MEXICO}

\email{jdo@astroscu.unam.mx, nunez@nucleares.unam.mx,
jzavala@nucleares.unam.mx}

\begin{abstract}
In this work we study the gravitational influence of the material
extending from Uranus orbit to the Kuiper belt and beyond on
objects moving within these regions. We conclude that a density
distribution given by $\rho(r)=\frac{1}{r}$ (for $r\geq 20 UA$)
generates a constant acceleration towards the Sun on those
objects, which, with the proper amount of mass, accounts for the
blue shift detected on the Pioneers space crafts. We also discuss
the effect of this gravitational pull on Neptune, and comment on
the possible origin of such a matter distribution.

\end{abstract}

\date{\today}

\maketitle

Mankind is now in direct contact with regions beyond the Solar
System. The space probes launched in the $70's$ already passed the orbit of
the last Solar system planet, Pluto, and are still working! It is really
a homage for their designers. In particular, the Pioneer probes 10 and
11 were designed in such a cunning way that their position, via the
Doppler effect, can be determined with great accuracy (the tracking system
have the sensitivity to measure frequency changes at the level of $mHz$/s, \cite{NTA1}).

When these probes were still within the Solar system, at 20 UA, that is,
between Uranus, which has a mean distance from the Sun of 19.13 UA, and
Neptune, at 30 UA, the frequency received at Earth start showing an
unaccounted
effect, a blue shift that is usually interpreted as a constant acceleration
with a magnitude of
\be a=8.74\pm1.33\,\times 10^{-8}\,\frac{\rm cm}{\rm s^2},
\label{pioneer}\ee
and directed towards the Sun \cite{NTA1}. The cause for such a
blue shift was unknown but more remarkable was the fact that such
phenomenon kept being present. In figure 1, we show
a diagram of the Solar system showing the Pioneers' trajectories and positions.
This figure is a reproduction taken from the original at \cite{drag}.
\begin{figure}[htb]
\includegraphics[width=7cm]{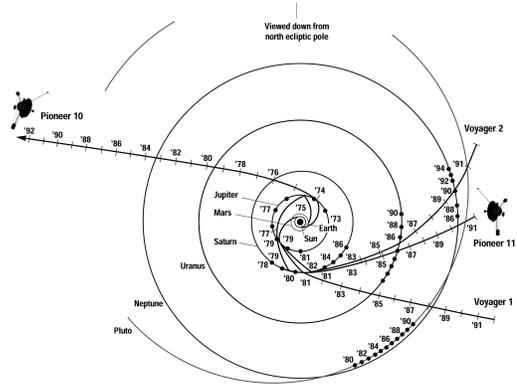}
\caption{\label{fig:tray} The Pioneers trajectories in the Solar
System \cite{drag}.}
\end{figure}

There have been a number of possible explanations for the cause of the
blue shift, and there is now an important number of works that discard any
internal effect of the probes as the cause, such
as heat reflected off the probes and possible gas leaks, (see \cite{NTA1},
\cite{drag} and \cite{NTA2}),
implying that it is due to an external cause.
Now the Pioneers are farther than $70$ UA, well beyond Pluto's orbit, $39.3$ UA,
and the effect is still there. Following Occam's razor, the simplest explanation is
that a {\bf constant} force, independent of the distance, is producing the measured blue shift.

In this way, the facts are that our first encounter with the external
regions of the Solar system and beyond are showing the presence of something
out there which affects the motion of
bodies. It is interesting to mention
that this effect started clearly as the probes passed Uranus orbit,
still within the Solar system, and it has not being observed, so far, in the trajectories
of the planets in that region, Neptune and Pluto.

Any number of explanations have being put forward to account for
this effect, including dark energy (see for example \cite{DE}),
quantum oscillations of the spacetime, branes (\cite{Branes}) and
you name it. (For other interesting alternatives see \cite{SQM},
\cite{vaquo} and \cite{clocks}). Our point of view is try first to
explain the phenomenon with local, everyday physics, and if this
is not enough, then use other alternatives. Also, it is clear that
dark energy contribution to the motion of bodies at local scales
is very much smaller than the one detected at the Pioneers
trajectories. Indeed, the effects of dark energy start to being
noticeable only at galaxy cluster scales! (see, for instance,
\cite{Peebles}). However, it is remarkable that, taking the
accepted value for the cosmological constant, considering it as
the source of the dark energy,
$\Lambda=3\,\Omega_{DE}\,\left(\frac{H_0}{c}\right)^2$, with
$\Omega_{DE}=0.7$, the ratio of dark energy density to the
critical density of the Universe, $H_0$ the Hubble constant today
(we take $h=0.7$ \cite{vh}), and $c$ the speed of light, we get
$\Lambda=1.2\,h\,\times\,10^{-56}\,{\rm cm}^{-2}$, and if we want
to construct an acceleration associated with it, we obtain that
$a_\Lambda=c^2\,\sqrt{\Lambda}=9.79\,\times\,10^{-8}\,\frac{\rm
cm}{\rm s^2}$, a value close to the one observed on the Pioneers
\cite{Smo}. Nevertheless, as mentioned, the cosmological effects
are negligible at Solar systems scales, so this is just a
remarkable coincidence.

In the present work we develop a more local and common idea. The Solar
system started from the proto-planetary cloud which, by gravitational collapse,
formed the Sun and planets but some material remained in the form of small
structures, tiny rocks, and dust, revolving around the Sun, forming belts.
Those tiny rocks within the orbits of the planets, were ultimately swept by them,
and the ones beyond form belts, two mayor regions beyond Neptune's orbit: The Kuiper belt, going from $30$ UA
to $70$ UA, where it joints with the \"Oort cloud, which extends up to 4000 UA
\cite{nubes,nubes1}, and internal belts within the orbits of the four major external planets.
We study the gravitational pull generated by those belts on the objects moving
within them and conclude that, if their density distribution goes as
$\rho(r)\propto\frac{1}{r}$, a constant acceleration pull is produced, directed towards the center,
which could account for the observed blue shift on the Pioneer probes.

As an illustration, let's take the
simplest case of a solid spherical distribution of matter, an
object moving within a media of a given density distribution
$\rho(\vec{r})$, will be accelerated according to the following
Newtonian law:
\be \vec{\nabla}\cdot\vec{a}=-4\,\pi\,G\,\rho(\vec{r}), \ee
with $G$ the gravitational constant,
$G=6.67\,\times\,10^{-8}\frac{{\rm cm}^3}{{\rm gr}\,s^2}$,
and $\vec{a}$ the acceleration of the object. Thus,
\be
\oint_S\,\vec{a}\cdot\hat{n}\,dA=-4\,\pi\,G\,\int_V\,\rho(\vec{r})d^3r, \ee
with $\hat{n}$ the normal vector to the surrounding surface to
the volume $V$. Considering that the acceleration is only radial:
$\vec{a}=a_P\,\hat{r}$, and taking a spherical shell of radius $r$,
as the surface of integration, we get
\be
a_P=-4\,\pi\,G\,\frac{1}{r^2}\,{\int_{0}}^{r}\rho(r')\,r'^2\,dr',
\label{eq:a_P}\ee
from which is clear that, in order to have a constant acceleration, the
density on the media must go as:
\be \rho(r)=\frac{\alpha}{r}\label{eq:rho} \ee
Moreover, we can compute the proportionality constant $\alpha$, from
the observed value of the Pioneers' acceleration; obtaining that:
\be \alpha=\frac{a_P}{2\,\pi\,G}= 0.21\,\frac{\rm gr}{{\rm
{cm}^2}}\ee

And then, the density of the sphere needed to produce
such an acceleration is, for example at $20$ UA where it's
claimed that the anomaly starts: $\rho(20UA)=7\times
10^{-16}\,\frac{\rm gr}{{\rm {cm}^3}}$. However, this solid
sphere is not the answer to the problem because it starts form
$r=0$, i.e. from the Sun, and the anomaly doesn't appear until
$20$ UA. The next possibility is a spherical shell starting at
$20$ UA with the same density distribution as in eq.
(\ref{eq:rho}). In this case the acceleration within the shell is
not constant: for the region inside $20$ UA the acceleration is
zero, for the region inside the shell, we can calculate the
acceleration by taking the difference between the influence of
two spheres centered in the Sun; the smaller one with a radius
$r_1=20$ UA and a mass $M_1$ produces a radial acceleration in a
test particle located at $r>20UA$ given by:
\be a_1=-\frac{GM_1}{r^2}=-\frac{G2\pi\alpha r_1^2 }{r^2}
\ee
The bigger sphere with radius $r$ produces an acceleration whose
magnitude is given by eq. (\ref{eq:a_P}): $a_2=-2\pi G\alpha$;
therefore, the acceleration inside the shell for a test particle
is: \be a_P=-2\pi G\alpha\left(1-\frac{r_1^2}{r^2}\right). \ee

\begin{figure}[htb]
\includegraphics[width=5cm]{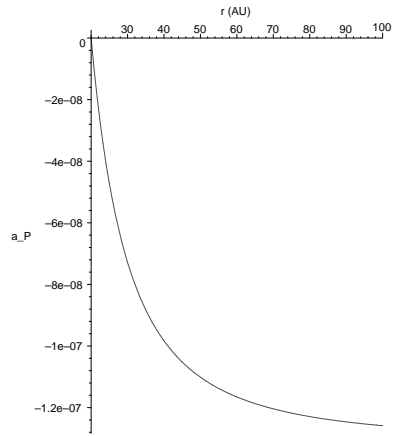}
\caption{\label{fig:shell} Deacceleration caused in a test
particle located at $r$ due to a spherical shell starting from
$r_1=20$ UA to $r_2=100$ UA, the acceleration is in units of
$1\times 10^{-8}\,\frac{\rm cm}{{\rm {s}^2}}$.}
\end{figure}

For this case the resulting acceleration is not
independent of the radius but as $r$ gets larger compared to
$r_1$ the acceleration approaches a constant value. If we fix
this value to eq. (\ref{pioneer}), then the value of $\alpha$ and
$\rho(20UA)$ are the same as in the case of the solid sphere.
Even when the acceleration is not constant within the sphere,
this case can't be discarded because of the uncertainty in the
value of the anomaly (around $15\%$, see eq. (\ref{pioneer}); see
also fig. 2 of \cite{NTA1}); from figure \ref{fig:shell} we can see that
for an important region inside the sphere the calculated acceleration
is in accordance with the
anomaly taking into account the uncertainty in the observational
data.

\begin{figure}[htb]
\includegraphics[width=7cm]{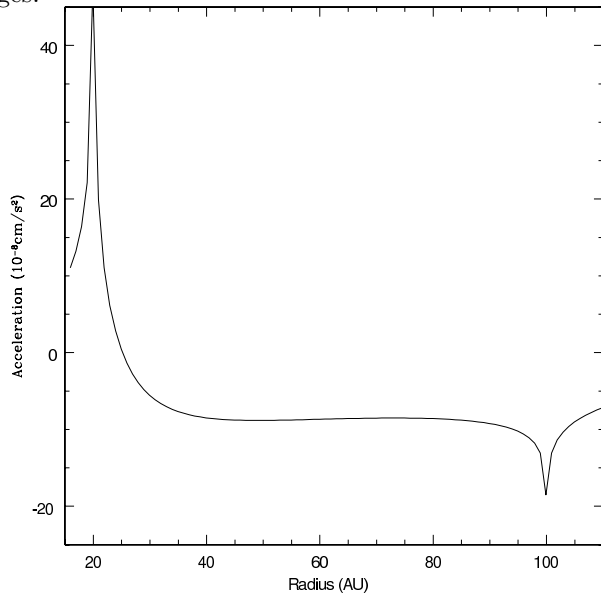}
\caption{\label{fig:disk} Acceleration caused in a test particle
located at $r$ due to an infinitely thin disk going from $r_1=20$
UA to $r_2=100$ UA, the acceleration is in units of $1\times
10^{-8}\,\frac{\rm cm}{{\rm {s}^2}}$.}
\end{figure}

The third case is a cylindrical ring, with height $h$ and
borders at $r_1=20$ UA and $r_2=100$ UA and a surface density
$\sigma(r)=\alpha_s/r$ (in this case, $r$ is the radius in
cylindrical coordinates), and the solution is not as
trivial as the two previous. In the appendix we describe the
method used to calculate the acceleration of an infinitely thin
disk ($h\rightarrow 0$), the result appears in figure
(\ref{fig:disk}). The following important features can be seen in
the figure: the acceleration is almost constant between $40$ UA
and $80$ UA with a change of less than $3\%$; near the edges of
the ring a significant change in the acceleration takes place,
this is due to the fact that $\vec{a}$ is singular at the edges.

The first feature of the results show us that a constant
acceleration can be produced in a large region inside the ring by
such a simple model. The region of this constancy can be extended
backwards (from $40$ UA to $20$ UA as the Pioneer anomaly
requires) by extending backwards the ring, it has to start then
at $r_1<20$ UA. However it's important to mention that in order
to give a better estimate of where this ring could start in
reality, it's necessary to have more precise data that the one
reported for the Pioneer anomaly, specially in the zone when the
anomaly starts to be significant, because it's in this region
where the larger uncertainty in the data is present (see figure 2
of \cite{NTA1}).

Regarding the second feature, we can say that the
singularities can disappear taking a smooth but fast change in the
surface density in the edges and not a cut-off as the one we used
in the calculations, this is is of course a more realistic case.
It's also important to say at this point that when the 2D case we
have chosen here is taken to 3D for a a cylindrical ring of
height $h$ the results doesn't change significantly, the
calculation has been made somewhere else \cite{Nf} (see figures 5
and 8 of the paper), for this case, in the region of constant
acceleration, the change in magnitude between the 2D and 3D
models is almost null, the main difference is near the edges
where the spikes become less extreme for the 3D case.

In order to obtain the magnitude we need for the
acceleration, the surface density at $r=20$ UA of the thin ring
must be, according to eq. (\ref{eq:a_P}) in the appendix:
\be \sigma(20UA)=0.81\frac{\rm gr}{{\rm {cm}^2}} \ee
In the realistic case of 3D this ring becomes a cylindrical ring
of height $h$ that has a volumetric density given by
$\rho=\sigma/h$, then for $r=20$ UA, and assuming $h=1$ UA we
have: $\rho(20UA)=5.4\times 10^{-14}\,\frac{\rm gr}{{\rm
{cm}^3}}$.

The total mass of a belt with such a density, considering a disk
with a thickness of $1$ UA, and ranging from $20$ to $100$ UA,
gives \textbf{$M_{disk}\approx 306 M_{\rm \oplus}$},
two order of magnitudes larger than the current estimates
on the Kuiper's belt mass \cite{peli}.

A first possible explanation for this discrepancy, which is
obtained considering that the belt is formed out of pure dust,
could be the presence of tiny ice rocks and gas which have not
been accounted for by those studies. In the same work,
\cite{peli}, the authors did consider, as in our present work, the
gravitational influence of the Kuiper belt as a possible
explanation to the acceleration seen on the Pioneers, but
discarded that influence based on two reasons: First that the
acceleration profile is not constant across the data range, (this
is because they did not consider a density distribution going as
$\frac{1}{r}$), and second that the total mass exceeded the
current estimations, $M_{disk}=0.3 M_{\rm Earth}$, for the dust in
the Kuiper belt region from $30$ to $100$ UA.

Nevertheless, even though such mass distribution does explain the
deacceleration observed in the Pioneers, it is unlikely that
baryonic matter could be responsible for this effect, there would
have to be a very high amount of small objects implying a large
collision rate and optical and infrared signatures that have not
been detected. Thus, in order to keep the model and mantain that
the observed deacceleration is due to gravitational pull, the
other possible candidate is the dark matter within the Solar
system. The fact that the galaxies, including ours of course, are
surrounded by dark matter halos is already a well established
fact. The numerical simulations, namely the NFW, predicts the dark
matter halo and give a well known density distribution \cite{NFW}.
Evaluating that density at the position of our Solar system, and
considering that the gravitational influence of the Sun extends to
the \"Oort cloud (around $7\times10^4$ UA), we see that there are
about $500 M_{\rm Earth}$ of dark matter whose dynamics is
dictated by the Sun. It is still an open question on how this
Solar dark matter halo reacts to that influence. When the proto
Solar system was being formed, the Solar dark matter halo was
pulled inwards and it could be that the final configuration is an
spherical shell or a belt in a region close to the Sun with
properties similar to the ones described in the last paragraphs,
giving in this way the mass needed to explain the observed
deacceleration on both spacecrafts.

\textbf{Given the properties requiered for the dark matter
particles, the natural candidates are WIMPs (Weakly Interacting
Massive Particles). Indeed, one most favoured WIMP to be the dark
matter particle is the neutralino, predicted by one of the
extensions of the standard model of particles, the Minimal
Supersymmetryic Standard Model (MSSM), which complies with the
characteristics needed for a dark matter particle, namely mass of
the order of $100 {\rm GeV}$, making it a member of the Cold Dark
Matter pradigm, very small cross section, $\equiv 10^{-9}\,{\rm
GeV}^{-2}$, electrically neutral and stable (\cite{Feng1,Feng2}).
Detection of the neutralino is nowadays an exciting field of
research, neutralinos may be detected either directly through
their interactions with ordinary matter or indirectly through
their annihilation decay products. In the first case, several
experiments on Earth are looking for signals associated with dark
matter scattering with baryonic matter (nucleus recoils for
example) (\cite{Dama,Cresst}) with no conclusive result yet. The
interpretation of this data depends on the spatial distribution of
density and velocity for dark mater in the Earth neighborhood
which is poorly known. The model presented here could serve as an
attempt to give a prediction for the density distribution of dark
matter using the data of the Pioneer's anomaly, but only for
regions beyond $20UA$, therefore it has no effect for direct
detection experiments on Earth. Indirect detection is based on
looking for signals of the remnants resulting from dark matter
annihilation, the neutralino being a Majorana particle can
annihilate with itself. The production of these remnants,
positrons, photons, neutrinos, etc., can be enhanced in regions of
high dark matter density such as the center of galaxies (for an
excellent review on this topic see \cite{Feng1}).}

\textbf{The properties of the neutralino fit very well with the
model presented in this work. Their presence influence the
dynamics of the objects mostly through gravitational interaction
and has a negligible dispersion with ordinary matter. And it is
for this last reason that there could not be a measurable
contribution to the Pionner anomaly due to dispersion effects.}

Also, it is important to point out that the belt would also
affect Neptune's orbit, modifying its period in the following way
(using the approximation of circular orbits):
\be
T^2=4\,\Pi^2\,\frac{{d_N}^3}{G\,M_\odot}\left(\frac{1}{1+\frac{a_P}{{d_N}^2}{G\,M_\odot}}\right),
\ee
that is, Neptune also feels the acceleration due to this matter
distribution.
Such acceleration changes Neptune's period in
$0.29$ seconds {\it per} period, implying a shift in Neptune's center of mass of $1.62$
kilometers after each revolution. It is a very tiny quantity but future probes could be
designed in such a way as to be able to measure these effects.

\begin{figure}[htb]
\includegraphics[width=7cm]{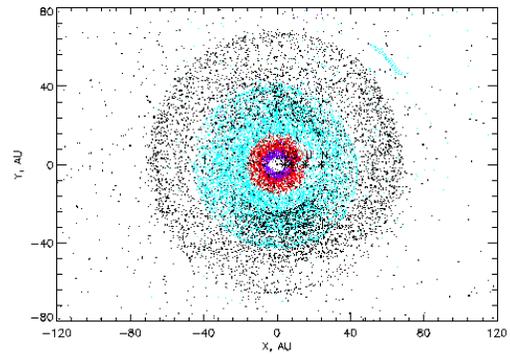}
\caption{\label{fig:kuiper}Numerical simulation of the density
distribution of matter at the Solar system in the ecliptic plane
\cite{Tania}.}
\end{figure}

\begin{figure}[htb]
\includegraphics[width=8cm]{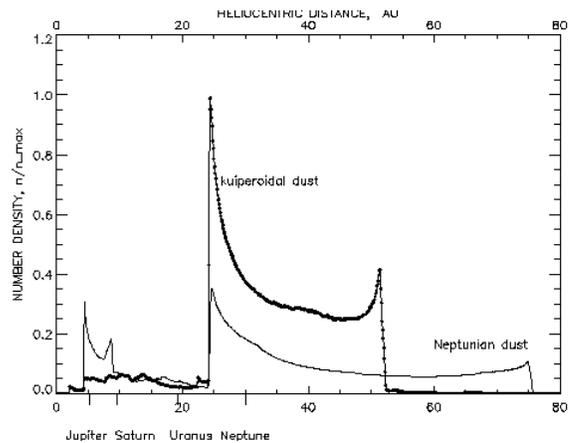}
\caption{\label{fig:densidad}Numerical simulation of the radial density
distribution of matter
at the Solar system in the ecliptic plane \cite{Tania}.}
\end{figure}

The radial density profile (going as $1/r$) needed to explain a constant
acceleration towards the Sun, can be explained by models of
Solar System formation (in the case of baryonic matter only). In \cite{Tania}, the authors
made an analysis of the dust population
in the outer solar system region; making a computational approach (see fig. \ref{fig:kuiper})
considering source objects from the Jupiter-family comets and the Kuiper Belt.
They obtained a radial distribution of dust for the region between 20 to 70 UA,
that has a sharp peak near 20 UA and from there falls like $\frac{1}{r}$, just as our model predicts.
After that, it has a smaller peak near 50 UA (see fig. \ref{fig:densidad}
of this paper that is a reproduction of fig. 5 of \cite{Tania}).
This result is encouraging for our model but given the large amount
of mass needed to reproduce the actual deacceleration measured by the
Pioneers, it would have to be remade with the inclusion of dark matter
to see if the density profile remains the same with the required amount of mass.


Following this scheme, the mass density distribution that appears
after the orbit of Uranus, is a result of an inward transport of
material from the Kuiper Belt into this orbit due mainly to the
gravitational influence of Neptune and Uranus. If the radial
distribution goes in fact like $\frac{1}{r}$ (at least between 20
to 100 UA), as the kind of models presented by Gor`kavyi {\it et
al.} \cite{Tania}, and with a total mass of $306M_{\oplus}$, then
the so called Pioneer anomaly can be explained by the
gravitational force that this distribution produce on every test
particle inside it, as was shown previously, making the observed
blue shift not an anomaly, but a natural consequence of the
gravitational pull due to the material present in these regions.

\textbf{Indeed, the actual dark matter distribution within the
Solar System is still an open question. A numerical simulation has
to be performed to give an appropriate description on this
subject, taking into account the effects of the formation and
evolution of the Solar system on the dark matter Solar halo. The
outcome of these analysis would determine whether or not the
matter distribution proposed in the present model is plausible. In
particular, it is important to determine the influence due to the
outer planets. This would answer the question if a gap of $20$ AU
starting from the Sun an then a region with density falling like
$1/r$ that has not been trashed away by Neptune is possible. The
gravitational influence of the outer planets on the baryonic
matter has formed Solar belts in these regions, which probably has
a matter distribution similar to the one required for dark
matter.}

Also, we want to comment that while this work was being finished, we
learned that Nieto {\it et. al} \cite{drag} were also considering the effect of
the material in the Kuiper belt on the space probes. However, in their approach
they studied the drag force of this material on the probes, obtaining
a different density distribution, mainly that the density should be constant in
order to explain the constant acceleration pull observed on the Pioneers.

Finally, we remark that it is very important to perform new and
more detailed studies on the density distribution and total mass
of the matter belts in the regions beyond Uranus' orbit to verify,
or discard, the explanation for the Pioneer acceleration presented
in this work. \textbf{In order to do so, we consider it is
necessary to send several probes to the outer regions of the Solar
System, like the TAU Probe (Thousand Astronomical Unit), in
several directions, including those perpendicular to the Solar
system plane, and equipped with the needed instruments in order to
determine, via the deacceleration effects, the nature and
characteristics of the phenomenon which causes these effects, and
then see if it is compatible with a dark matter distribution.}

We acknowledge that the required amount of mass needed to reproduce the
magnitude of the Pioneers deacceleration is indeed very large considering
the actual estimates of mass in such a region, however the idea proposed
in this work is appealing due to it's simplicity and has already generated
some attention in the scientific community (see for example the works that
have appeared while this work was being considered for publication,
\cite{Nf,Bertolami}). We think that the proposal described above considering
dark matter as a possible explanation for the unaccounted mass is worth to
follow and, as we mentioned above, it would be interesting to perform numerical simulations for the
Solar System formation which include dark matter in order to study its final distribution.

This work was partially supported by the grants DGAPA-UNAM
IN113002, IN122002. JZ also acknowledges the scholarship support
by CONACyT and DGEP. We want to thank our colleagues in the
Gravitational department of the Nuclear Science Institute for much
encouragement and helpful discussions on the present idea. We also
thank the anonymous referee of the present work for several
remarks which contributed to better sketch the problems and
advantages of the model proposed.

\section{Appendix}

\bigskip

In the present appendix we will describe the procedure to obtain
the exact result for the acceleration caused by a distribution of
matter in a infinitely thin ring going from $r_1=20$ UA to
$r_2=100$ UA with a surface density distribution:
$\sigma(\vec{r})=\alpha_s/r$.

The method is straightforward, using Newton's law of gravitation,
we calculate the acceleration caused by the distribution of
matter on a test particle; the case of interest for us, is only
when the position of the test particle (the Pioneers) is on the
plane of the disk, therefore we need only two dimensions to do
the calculations.

An infinitesimal element of mass $dm_i$ and area $ds_i$, produces
an acceleration $d\vec{a_i}$ in the test particle placed in the
position $\vec{r}$ given by:

\be
d\vec{a_i}(\vec{r})=-\frac{Gdm_i(\vec{r}-\vec{r_i})}{|\vec{r}-\vec{r_i}|^3}
=-\frac{G\sigma(\vec{r_i})ds_i(\vec{r}-\vec{r_i})}{|\vec{r}-\vec{r_i}|^3}
\ee

\noindent where $\sigma(\vec{r_i})$ is the surface density of the mass
element. Then all we have to do is to sum the contribution of all
the mass elements in the ring. To do so, we use the following
simplifications: if we assume that
$\sigma(\vec{r_i})=\sigma(r_i)$, then the symmetry of the problem
allow us to put, choosing a cartesian coordinate system,
$\vec{r}=x\hat{e_x}$ and
$\vec{r_i}=r_i(cos\theta_i\hat{e_x}+sin\theta_i\hat{e_y})$, where
$r_i$ and $\theta_i$ are the coordinates of the mass element
$dm_i$ in polar coordinates. The total contribution to the
acceleration in the direction of $\hat{e_y}$ is zero given the
symmetry of the ring; for the component in the direction
$\hat{e_x}$ the integral over the whole distribution ($\theta$
going from $0$ to $2\pi$ and $r_i$ going from $r_1$ to $r_2$)
gives the radial acceleration caused on the test particle as a
function of it's radial position. To perform these calculations
we made a numerical program in fortran 77. The resulting
acceleration as a function of the position $r$ of the test
particle, appears in fig. (\ref{fig:kuiper}) of the paper, the
value of $\alpha_s$ was adjusted to obtain the magnitude of the
acceleration in the Pioneer anomaly, the value is:
\be \label{alfa_s} \alpha_s=2.43\times 10^{14}\,\frac{\rm gr}{{\rm
cm}} \ee

\end{document}